\pdfoutput=1
\documentclass[12pt,a4paper]{article}
\usepackage[utf8]{inputenc}
\usepackage[T1]{fontenc}
\usepackage{amsmath,amsfonts,amssymb}
\usepackage{graphicx} 
\usepackage{booktabs}
\usepackage{algorithm}
\usepackage{algorithmic}
\usepackage[margin=1in]{geometry}
\usepackage{url}
\usepackage{float}

\title{Research on Evaluation Methods for Patent Novelty Search Systems and Empirical Analysis}
\author{Shu Zhang, LiSha Zhang, Kai Duan, XinKai Sun}
\date{August 2025}

\begin{document}

\maketitle

\begin{abstract} 
\itshape Patent novelty search systems are critical to IP protection and innovation assessment; their retrieval accuracy directly impacts patent quality. We propose a comprehensive evaluation methodology that builds high-quality, reproducible datasets from examiner citations and X-type citations extracted from technically consistent family patents, and evaluates systems using invention descriptions as inputs. Using Top-k Detection Rate and Recall as core metrics, we further conduct multi-dimensional analyses by language, technical field (IPC), and filing jurisdiction. Experiments show the method effectively exposes performance differences across scenarios and offers actionable evidence for system improvement. The framework is scalable and practical, providing a useful reference for development and optimization of patent novelty search systems.
\end{abstract}

\section{Introduction} 
Global patent filings continue to grow, and patent novelty search systems are pivotal for assessing the novelty and inventiveness of technical solutions. Unlike general text retrieval, patent documents are long, structurally complex (abstract, claims, description), terminology-dense, and legally consequential, which raises the bar for text understanding, semantic matching, and ranking.

However, current evaluations of novelty search systems face three common limitations: (1) the lack of unified benchmarks and high-quality datasets, making cross-system comparison difficult; (2) heterogeneous evaluation processes and metrics that emphasize aggregate scores while lacking fine-grained analysis across languages, technical fields, and countries; and (3) heavy reliance on manual annotation, which is costly and prone to subjective bias. These issues hinder objective assessment and sustained improvement of novelty search systems.

To address this, we propose a comprehensive evaluation methodology that leverages examiner citation relationships as objective relevance signals and augments them with X-type cited references from technically consistent family patents to construct automated, reproducible evaluation data. Using the invention description as input, we assess systems with Top-k detection rate and recall, and conduct multi-dimensional analyses by language, technical field (IPC), and filing country to reveal performance characteristics under varied scenarios.

The main contributions are: 
\begin{enumerate} 
\item An automated dataset construction method based on examiner citations and family-patent X-type citations; 
\item A standardized evaluation process using invention descriptions and core metrics (Top-k detection rate, recall); 
\item A multi-dimensional analysis framework covering language, technical field, and filing country; 
\item Large-scale empirical validation demonstrating the method’s effectiveness and scalability. 
\end{enumerate}

\section{Related Work} 
Evaluation methods for patent search systems are mainly divided into the following categories:

Manual annotation-based evaluation methods construct test sets through expert annotation, but suffer from high annotation costs and difficulty in ensuring consistency across annotators.

User behavior-based evaluation methods leverage click data and user feedback for assessment, yet face challenges related to data sparsity and various forms of user bias.

Patent citation relationship-based evaluation methods utilize citation networks between patents (e.g., examiner citations in search reports) to construct evaluation datasets, providing a degree of objectivity but limited in coverage and potentially affected by citation practices.

Benchmark dataset approaches rely on established collections and shared tasks such as NTCIR Patent Retrieval and CLEF-IP, but these often fail to capture emerging technological domains and real-world search scenarios.

\section{Evaluation Methodology}
\subsection{Dataset Construction} 
The dataset construction method proposed in this paper is based on the inherent structural characteristics of patent literature, utilizing patent cited references and comparative documents (x) from family patents with identical claim protection scope to construct high-quality evaluation datasets.

\begin{figure}[H]
\centering
\includegraphics[width=0.8\textwidth]{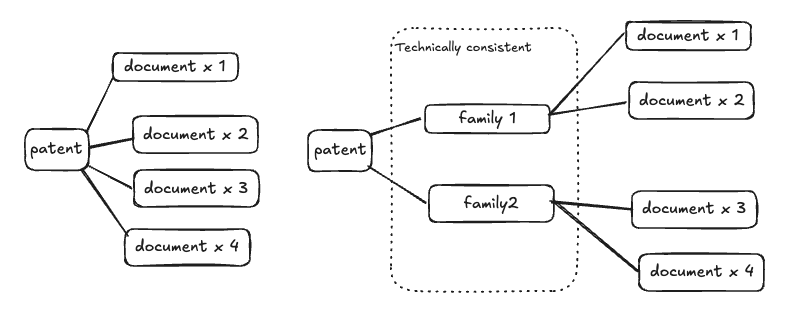}
\caption{\label{fig:family_patents} document x relationships}
\end{figure}

Figure \ref{fig:family_patents} shows the relationship between main patent and document x.

\begin{figure}[H]
\centering
\includegraphics[width=0.8\textwidth]{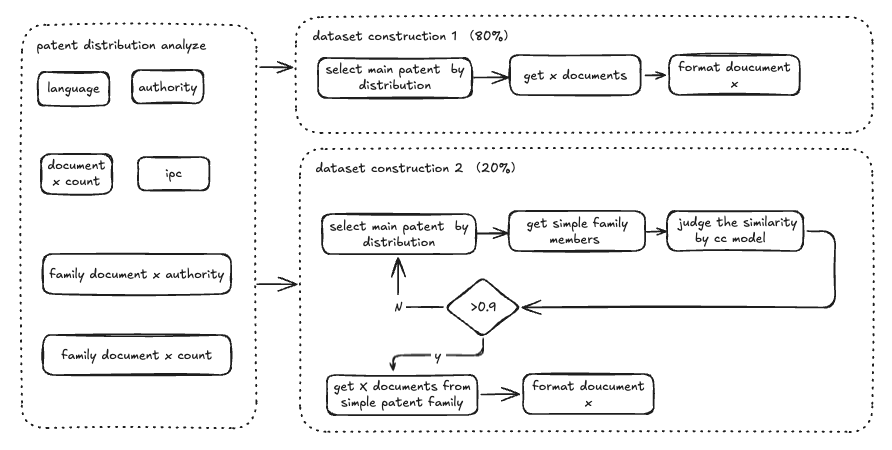}
\caption{\label{fig:dataset construction} Dataset construction}
\end{figure}

Figure \ref{fig:dataset construction} shows the dataset construction process

\subsubsection{Utilization of Patent Cited References} 
Patent cited references are documents identified during the examination process that represent prior art relevant to a patent application. These references, particularly X-type citations (indicating documents that challenge novelty) and Y-type citations (indicating documents that challenge inventive step when combined with other references), provide valuable technical comparisons for patent assessment. The distribution of these citation types in our dataset reflects their actual occurrence in patent examination processes, ensuring a realistic evaluation framework.

The cited reference extraction process is as follows: 
\begin{enumerate} 
\item Analyze the entire patent corpus to identify patents that contain comparative document X, and calculate the distribution proportion of patents where the citation type is X
\item Analyze language distribution of both primary patents and comparative documents (X-type references) to ensure the dataset reflects the actual language diversity in global patent filings; 
\item Analyze the geographical distribution of primary patents to ensure it reflects the actual global patent filing patterns across different countries and regions; 
\item Prepare and organize the dataset according to the actual distribution ratios to ensure the dataset reflects real-world citation patterns and language distributions. 
\end{enumerate}

\subsubsection{Family Patent Identification and Utilization} 
Family patents refer to patents filed in different countries or regions based on the same or related inventions, which have high similarity in technical content. Instead of directly using family patents, we utilized the X-type cited reference lists corresponding to the family patents of the main patent. This approach considers: 
\begin{enumerate} 
\item Country distribution of family patents: Ensuring geographical representation; 
\item Technical alignment: We evaluated alignment among members of the same patent family using an in-house lightweight CC model trained with reinforcement learning to assess technical similarity between patent documents. The model generates a technical-similarity score for each member, and those with CC scores of 0.90 or higher were retained. Extensive validation indicates that this threshold reliably ensures technical alignment and helps prevent the inclusion of technically truncated counterparts that can arise across jurisdictions; 

\item Extraction of relevant X-type cited references: From technically consistent family patents to form a comprehensive set of similar technical solutions. 
\end{enumerate}

This methodology allows us to obtain lists of similar technical solutions that are relevant to the main patent through its family patents, creating a more robust and technologically representative dataset.

\subsubsection{Dataset Quality Control} 
To ensure evaluation dataset quality, the following rigorous quality control measures were further implemented: 
\begin{enumerate} 
\item Technical relevance assessment: Enhanced the cc model through reinforcement learning with extensive comparative document pairs, ensuring highly accurate determination of technical similarity between patents and their cited references. 
\item Patent recency criteria: Deliberately selected patents filed within the last decade to eliminate evaluation bias caused by differences in technical description styles between older and more recent patent documents. 
\item Data integrity verification: Conducted comprehensive systematic checks to ensure complete integrity and usability of all patent documentation, including proper formatting, presence of all required sections, and thorough validation of citation links. 
\end{enumerate}

\subsection{Evaluation Process Design}
\subsubsection{Input Processing} 
The evaluation process uses the invention description section of patents as technical input to the novelty search system. The invention description is the most detailed and complete part of technical content in patent documents, containing background, objectives, overview of implementation methods, and other information of the technical solution.

\begin{figure}[H]
\centering
\includegraphics[width=0.8\textwidth]{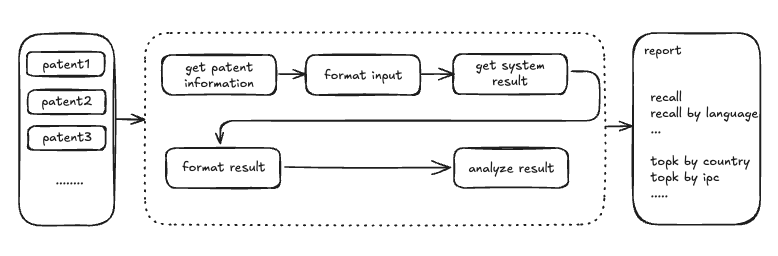}
\caption{\label{fig:evaluation process} Evaluation process}
\end{figure}

Figure \ref{fig:evaluation process} shows the evaluation process

Input processing includes the following steps: 
\begin{enumerate} 
\item Text preprocessing: Removing format markers, unifying encoding formats; 
\item Structural parsing: Identifying various components of the invention description; 
\item Key information extraction: Extracting and identifying the background and detailed description sections of the specification; 
\item Query construction: Converting processed technical content into query formats acceptable to the system. 
\end{enumerate}

\subsubsection{System Execution and Result Collection}
We submit the constructed queries—derived from the invention-description section—to the system under evaluation and collect the returned result lists. To ensure evaluation fairness and objectivity, the following principles are adopted: 
\begin{enumerate} 
\item Unified query interface and controls: Consistent API endpoints, random seeds, time limits, and maximum result depth across systems;; 
\item Result standardization: Converting output results from different systems into unified formats; 
\item Result recording: Completely recording query processes and system response information. 
\end{enumerate}

\subsection{Evaluation Metrics}
\subsubsection{Detection Rate (Top-k)}
Detection rate measures the probability of finding at least one relevant patent within the top-k positions of system-returned results, defined as:

\begin{equation} 
\text{Top-}k \text{ Detection Rate} = \frac{\left|\{\, q \in Q \mid \text{Rel}(q) \cap \text{Top}_k(q) eq \emptyset \,\}\right|}{|Q|}
\end{equation}
where $Q$ is the query set, $\text{Rel}(q)$ the set of relevant patents (examiner/family X-type), $\text{Top}_k(q)$ the top-$k$ results. This metric has a cumulative property: if a relevant document appears in a higher position (e.g., top-1), it will also be counted in lower positions (e.g., top-3, top-5, etc.). Therefore, Top-$k$ Detection Rate is monotonically non-decreasing as $k$ increases. Typically, we evaluate at $k = 1, 3, 5, 10$, and other selected values to understand retrieval performance at different depths.
\subsubsection{Recall} 
Recall measures the proportion of relevant patents retrieved by the system out of all relevant patents, defined as:
\begin{equation} 
\text{Recall} = \frac{|{\text{relevant patents}} \cap {\text{retrieved patents}}|}{|{\text{relevant patents}}|} 
\end{equation}

\subsection{Multi-dimensional Extended Analysis}
To deeply analyze novelty search system performance under different scenarios, this paper designs a multi-dimensional extended analysis framework.

\subsubsection{Analysis in Language Dimension}
Patent documents in different languages exhibit significant differences in expression styles, terminology usage, etc. The analysis in language dimension includes: 
\begin{enumerate} 
\item Monolingual retrieval performance: Evaluating the retrieval accuracy of the system within Chinese, English, and other language texts; 
\item recall capability: Assessing the system's ability to retrieve patents in languages different from the input language; 
\item X-type recall: Assess the ability to retrieve X-type references written in a language different from the primary patent language.
\end{enumerate}

\subsubsection{Technical Field Dimension Analysis}
Patents in different technical fields differ in technical complexity, terminology density, expression standards, etc. Technical field analysis includes: 
\begin{enumerate} 
\item Using IPC (International Patent Classification) for technical field classification; comparing system retrieval performance across different technical fields; 
\end{enumerate}

\subsubsection{Country Dimension Analysis}
Patent applications from different authorities differ in examination standards, document formats, language expression, etc. Country dimension analysis primarily reflects the retrieval accuracy of comparative patent literature across different countries, allowing evaluation of how well the model performs when handling patents with varying international standards and conventions.

\section{Experimental Design and Result Analysis}

\subsection{Experimental Environment and Data}

\subsubsection{Dataset Construction}
The evaluation dataset constructed in this study contains the following scale: 
\begin{itemize} 
\item Patent sample quantity: 10,00 patents 
\item Technical field coverage: Including 8 major technical fields such as electronics and communication, mechanical engineering, chemical pharmaceuticals, biotechnology 
\item Country distribution: 5,00 Chinese patents, 2,00 US patents, 2,00 European patents, 1,00 WO patents 
\item Average cited reference quantity: Each patent contains an average of 2 cited references 
\item Family patent coverage: 20\% of patents have family patent cited references  
\end{itemize}

\subsubsection{Evaluation System}
\begin{enumerate} 
\item PatSnap Semantic Search (Baseline) 
\item PatSnap Novelty Search (AI-based) 
\end{enumerate}

\subsection{Overall Performance Analysis}
\subsubsection{Main Metric Performance}
Table \ref{tab:overall_performance} shows the overall performance of the two systems.

\begin{center}
\label{tab:overall_performance}
\begin{tabular}{||c c c c c c c c c c||} 
 \hline
 System & Top1 & Top3 & Top5 & Top10 & Top20 & Top30 & Top50 & Top100 & recall \\ [0.5ex] 
 \hline\hline
 Semantic Search & 11\% & 16\% & 20\% & 24\% & 29\% & 33\% & 38\% & 44\%  & 0.32 \\ 
 \hline
 Novelty Search & 17\% & 26\% & 31\% & 39\% & 46\% & 53\% & 59\% & 67\% & 0.43 \\ [1ex] 
 \hline
\end{tabular}
\end{center}

Results show that Novelty Search outperforms Semantic Search across all metrics; improvements in Top 10 Detection Rate (+15 pp) and Recall (+0.11 absolute) are statistically significant under stratified paired bootstrap (p<0.01).

This evaluation methodology effectively identifies system limitations by measuring performance at multiple retrieval depths. This approach reveals where each system reaches diminishing returns and distinguishes between ranking problems and overall retrieval capability. The gap between Top 100 performance and recall highlights specific areas for improvement in both systems, providing clear direction for future enhancements in patent search technology.

\subsection{Multi-dimensional Performance Analysis}

\subsubsection{Language Dimension Analysis Results}

\begin{figure}[H]
\centering
\includegraphics[width=0.8\textwidth]{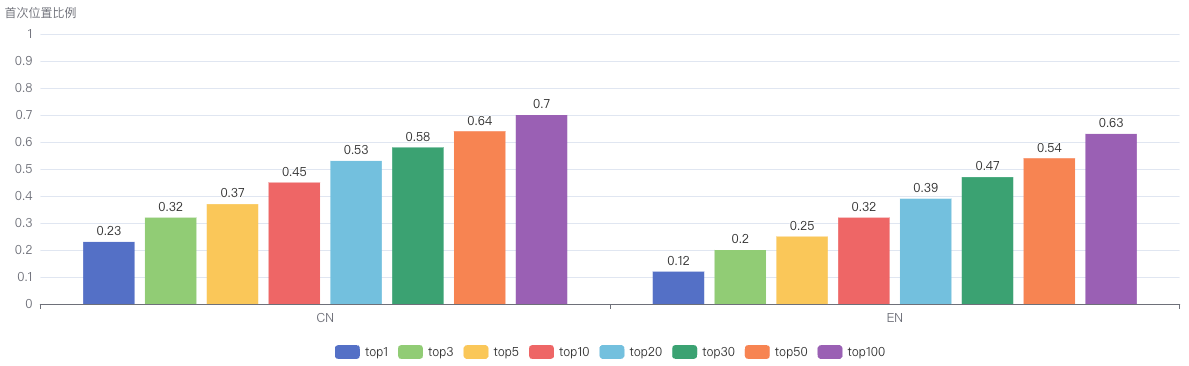}
\caption{\label{fig:language_analysis} language vs topk}
\end{figure}

Figure \ref{fig:language_analysis} shows the performance of PatSnap novelty search across language dimensions.

Analysis results indicate: 
\begin{enumerate} 
\item Chinese inputs consistently outperform English inputs across all ranking thresholds; 
\item At top-1, Chinese achieves 23\% accuracy vs 12\% for English (nearly double); 
\item The performance gap narrows at higher ranks (top-100: CN 70\% vs EN 63\%).
\end{enumerate}

This evaluation methodology effectively pinpoints language-dependent performance gaps across multiple ranking thresholds, enabling quantification of cross-language retrieval differences and prioritization of optimization efforts for weaker language inputs, prioritize optimization efforts for weaker language inputs, and make data-driven decisions about acceptable performance levels.

\subsubsection{Technical Field Dimension Analysis Results}

\begin{figure}[H]
\centering
\includegraphics[width=0.8\textwidth]{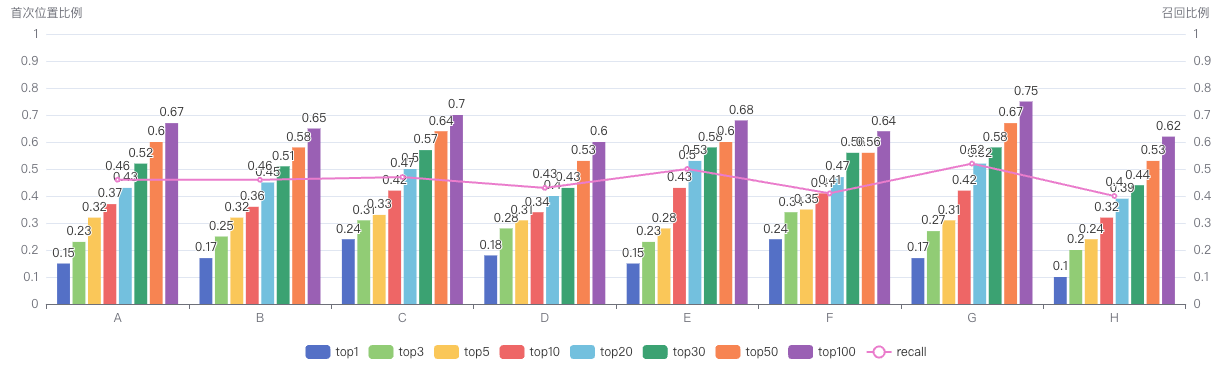}
\caption{\label{fig:ipc_analysis} ipc vs topk}
\end{figure}

Figure \ref{fig:ipc_analysis} shows the performance of PatSnap novelty search  across different technical fields.

Observations reveal: 
\begin{enumerate} 
\item Classification G shows the best overall performance (75\% at top-100) in this dataset.
\item Classification H shows the weakest top-1 performance (10\%), likely due to dense abbreviations, circuit notations, and diagram-heavy disclosures; Classification C can be challenging due to chemical formulae and specialized terminology.
\item Classifications C and F perform relatively well at top-1 (24\% each)
\item Recall rates vary significantly: highest in G (52\%), lowest in H and F (40-41\%).
\item Performance differences between classes are more pronounced at lower rank positions.
\item Even in the best-performing category (G), 25\% of relevant documents aren't found in top-100
\end{enumerate}

This evaluation approach effectively reveals technology domain-specific performance variations, enabling targeted optimization of search algorithms for underperforming patent classifications while highlighting where domain expertise or specialized models might be required. The combined ranking and recall metrics provide comprehensive insight into both precision and coverage issues across different technical fields, allowing for data-driven prioritization of improvement efforts.

\subsubsection{Country Dimension Analysis Results}

\begin{figure}[H]
\centering
\includegraphics[width=0.8\textwidth]{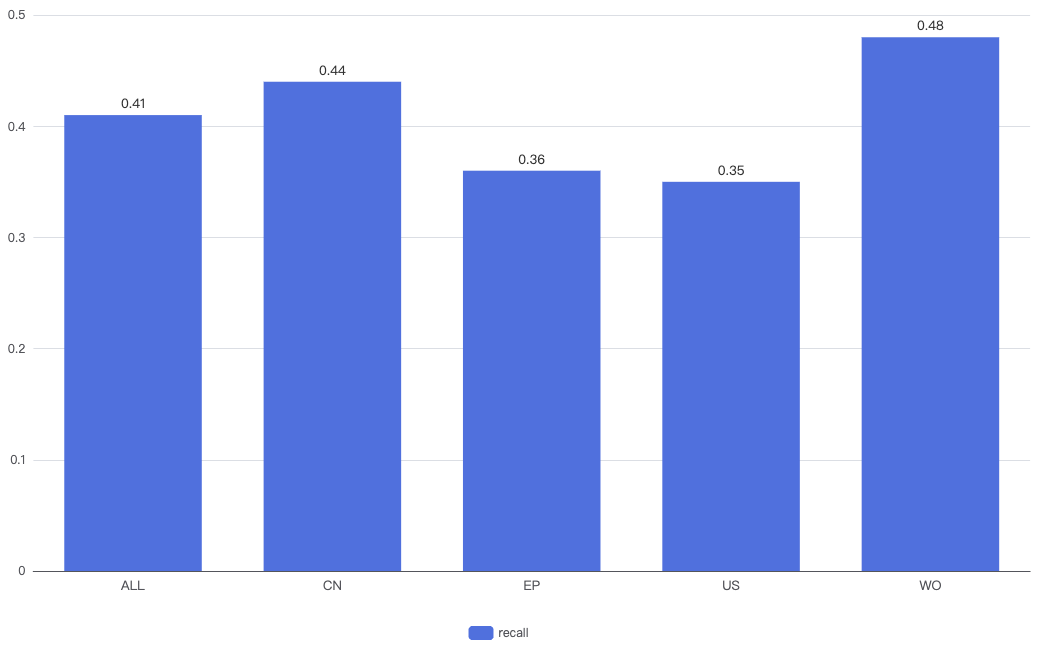}
\caption{\label{fig:country_analysis} country vs recall}
\end{figure}

Figure \ref{fig:country_analysis} shows the performance of PatSnap novelty search  across different authorities.

Analysis of retrieval performance across different country patents reveals the following patterns:

\begin{enumerate} 
\item WO authority (WIPO) shows the highest recall at 48\%; 
\item CN applications show above-average recall at 44\%; 
\item EP and US applications have notably lower recall rates (36\% and 35\% respectively); 
\item The 13\% gap between highest and lowest recall rates is significant. 
\end{enumerate}

\section{Conclusion}
This paper proposes a comprehensive evaluation methodology for patent novelty search systems. By constructing high-quality evaluation datasets and designing multi-dimensional analysis frameworks, it provides scientific methods for objectively evaluating novelty search system performance. Experimental results demonstrate that the proposed evaluation method can effectively identify system performance differences across different dimensions, providing valuable guidance for system improvement.

\end{document}